\documentclass[aps,prl,reprint,showpacs]{revtex4-1}
\usepackage{graphicx}
\usepackage{dcolumn}
\usepackage{bm}

\begin{document}

\title{The light filament as vector solitary wave}

\author{Lubomir M. Kovachev}
\affiliation{Institute of Electronics, Bulgarian Academy of Sciences,\\
Tzarigradcko shossee 72,1784 Sofia, Bulgaria}
\email[]{lubomirkovach@yahoo.com}
\date{\today}

\begin{abstract}
We present an analytical approach to the theory of nonlinear
propagation  of femtosecond optical pulses with broad-band spectrum
in gases. The vector character of the nonlinear third-order
polarization is investigated in details, taking into account the
carrier to envelope phase. The corresponding system of vector
amplitude equations is written by using left-hand and right-hand
circular components of the electrical field. We found that this
system nonlinear equations admits $3D+1$ vector soliton solution
with Lorentz shape. The solution presents relatively stable propagation
and rotation with GHz frequency of the vector of the electrical
field in plane, orthogonal to the direction of propagation. The
evolution of the intensity profile demonstrate weak self-compression
and week spherical wave in the first milliseconds of  propagation.
\end{abstract}

\maketitle

\section{Introduction}

When a femtosecond  laser pulse with power above the critical for
self-focusing propagates in air, a number of new physical effects
have been observed, such as long-range self-channeling
\cite{MOU,WO}, coherent and incoherent GHz and THz emission
\cite{TZ, DAMYZ}, asymmetric pulse shaping, super-broad spectra
\cite{HAU, CHIN1}, self-compression \cite{KOPR}, polarization
rotation \cite{KOZA} and others. A remarkable effect is also that
some of the light pulses propagate over distances of several
kilometers in vertical traces, preserving their spectrum and shapes
\cite{WO}. In one typical experiment in the near zone up to $1-3$
$m$ from the source, when the pulse's intensity exceeds $I>10^{12}
W/cm^2$, initial self-focusing and self-compressing starts, which
leads to enlarging the $k_z$ spectrum to broad band asymmetric one
$\triangle k_z\approx k_0$. The process increases the core intensity
up to $10^{14} W/cm^2$, where a short non-homogenous plasma column
in the nonlinear focus is observed. Usually the standard model
describing the propagation in the near zone is a scalar
spatio-temporal nonlinear paraxial equation including in addition
terms with plasma ionization, higher order Kerr terms, multiphoton
ionization and others \cite{KAN, KASP2}. The basic model works
properly in the near zone because of the fact, that paraxial
approximation is valid for pulses with narrow-band spectrum
$\triangle k_z << k_0$. In the far-away zone plasma generation and
higher-order Kerr terms are also included as necessary for the
balance between the self-focussing and plasma defocussing and for
obtaining long range self-channeling in gases. However, the above
explanation of filamentation is difficult to be applied in the
far-away zone. As reviewed in \cite{MECH,MECH2, KASP2, KILO} the
plasma density at long distances from the source is to small to
prevent self-focusing. There are basically three main
characteristics which remain the same at these distances - the
broad-band spectrum, coherent GHz generation and the width of the
core, while the intensity in a stable filament drops to a value of
$10^{12} W/cm^2$. The higher-order Kerr terms for pulses with
intensities of order of $I\sim 10^{12} W/cm^2$  are also too small
to prevent self-focussing. The experiments, where observation of
long-range self-channeling without ionization was realized
\cite{MECH, MECH2,KILO}, show the needing to change the role of the
plasma in the laser filamentation. In addition, there are
difficulties with the physical interpretation of the coherent GHz
radiation as a result of plasma generation. The light pulse near the
nonlinear focus emits incoherent and \emph{non-homogenous} plasma,
while the coherent GHz radiation requires \emph{homogenous} plasma
with fixed electron density of the order of $10^{15}$ $cm^{-3}$.
Only homogenous plasma can generate coherent GHz emission, but such
kind of plasma absent in the process of filamentation. In the
experiments the following basic characteristic of the long-range
filaments in air are detected:

1. Broad-band spectrum ($\Delta k_z \sim k_0$).

2. Intensity a little above of the value for self-focusing ($I\cong
I_{self-foc} \simeq 10^{12}$ $W/cm^2$).

3. Absence of plasma at long distances.

4. Asymmetric relatively stable (Lorentz) spectral and longitudinal
forms.

5. Coherent GHz generation.

Recently in \cite{KOVBOOK} was developed scalar ionization-free
non-paraxial nonlinear model which gives the above characteristics
of the stable filament. The analytical and the numerical results
describes correctly the linear and nonlinear evolution of
narrow-band and broad-band laser pulses and in addition it was
found, that the equation admits exact Lorentz-type soliton solutions
in approximation of neglecting the GHz oscillation. Nevertheless
this theory encountered significant difficulties. The main problems
are:

1. Peak instability of the soliton solution under small initial
perturbations.

2.  The soliton solution is obtained after neglecting the GHz
oscillation.

3. The soliton solution admits one free parameter.

4. There are problems with the conservation law of the nonlinear
operator, when we use the GHz oscillation.

To solve the above problems  we propose in this paper a nonlinear
vector model.

\section {Nonlinear Polarization}

The self-action process broadens the pulse spectrum - starting with
narrow-band pulse, the stable filament becomes broad-band far way
from the source. From other hand, the standard filamentation model
based on plasma generation and multi-photon processes includes also
nonlinear polarization of the kind

\begin{eqnarray}
\label{NLKERR} \vec{P}^{nl} = n_2 \left[\left( \vec{E} \cdot
\vec{E}^* \right)\vec{E} + \frac{1}{2} \left( \vec{E} \cdot \vec{E}
\right)\vec{E}^*\right],
\end{eqnarray}
where $n_2$ is the nonlinear refractive index of the isotropic
media.  The polarization (\ref{NLKERR}) was proposed by Maker and
Terhune in 1965  \cite{MAKER}. If the electrical field contains one
linear or circular component, the polarization (\ref{NLKERR})
describes only the self-action effect, while in the case of
two-component electrical filed $\vec{E} = (E_x, E_y, 0)$ additional
terms appear, presenting cross-modulation and degenerate four-photon
parametric mixing. The self-action process broadens the pulse
spectrum - starting from narrow-band pulse, the stable filament
becomes broad-band far way from the source. Later in
\cite{KOL1,KOL2,LMK1} it is shown, that the  evolution of broad-band
pulses like filaments can not be described correctly by nonlinear
polarization of the kind (\ref{NLKERR}). It is more correct to use
the generalized nonlinear operator

\begin{eqnarray}
\label{NLTH} \vec{P}^{nl} = n_2 \left( \vec{E} \cdot \vec{E}
\right)\vec{E},
\end{eqnarray}
which includes additional processes associated with third harmonic
generation (THG). The more precise analysis, presented in the paper,
demonstrates that the polarization of kind (\ref{NLTH}) is not
applicable to a scalar model, because the corresponding Manley-Rowe
(MR) conservation laws are not satisfied. That is why we substitute
into the nonlinear operators (\ref{NLKERR}) and (\ref{NLTH})
two-component electrical vector field at one carrying frequency in
the form:

\begin{eqnarray}
\label{ELF} \vec{E} =\frac{ \left( A_x
\exp\left[ik_0(z-v_{ph}t)\right] + c.c. \right)}{2}\vec{x} +\nonumber\\
 \frac{\left( A_y \exp\left[ik_0(z-v_{ph}t)\right] - c.c.
\right)}{2i}\vec{y},
\end{eqnarray}
where $A_x=A_x(x, y, z, t), A_y = A_y(x, y, z, t)$ are the amplitude
functions,  $k_0$ is carrying wave number, $\omega_0$ is the
carrying frequency of the laser source and $v_{ph}$ is the phase
velocity. Here we take in mind the fact that the $x$ and $y$
components are orthogonal and complex conjugated. In the case of
Maker and Terhune polarization (\ref{NLKERR}) we obtain for $A_x$
and $A_y$ components the following expressions:

\begin{eqnarray}
\label{POLKERR} \vec{P}^{nl}_x = \frac{3}{8} n_2 \left[ \left(
|A_x|^2 + \frac{2}{3}|A_y|^2\right) A_x - \frac{1}{3} A_x^* A_y^2
\right]\times \nonumber\\
\times\exp\left[ik_0(z-v_{ph}t)\right] + c.c.\nonumber\\
\\
\vec{P}^{nl}_y = \frac{3}{8i} n_2 \left[ \left( |A_y|^2 +
\frac{2}{3}|A_x|^2\right) A_y - \frac{1}{3} A_y^* A_x^2
\right]\times\nonumber\\
\times\exp\left[ik_0(z-v_{ph}t)\right] + c.c. \nonumber
\end{eqnarray}
The nonrestricted nonlinear polarization (\ref{NLTH}) generates the
following components

\begin{eqnarray}
\label{POLTH} \vec{P}^{nl}_x = \tilde{n}_2 \Bigg[ \frac{1}{3} \left(
A_x^2 - A_y^2\right) A_x \exp\left[2ik_0(z-v_{ph}t)\right]\nonumber\\
+ \left( |A_x|^2 +\frac{2}{3}|A_y|^2\right) A_x
- \frac{1}{3} A_x^* A_y^2
\Bigg]\exp\left[ik_0(z-v_{ph}t)\right] + c.c. \nonumber\\
\\
\vec{P}^{nl}_y = \frac{\tilde{n}_2 }{i} \Bigg[ \frac{1}{3} \left(
A_x^2 - A_y^2\right) A_y \exp\left[2ik_0(z-v_{ph}t)\right]\nonumber\\
+ \left( |A_y|^2 +\frac{2}{3}|A_x|^2\right) A_y
- \frac{1}{3} A_y^* A_x^2
\Bigg]\exp\left[ik_0(z-v_{ph}t)\right] + c.c., \nonumber
\end{eqnarray}
where $\tilde{n}_2= \frac{3}{8} n_2 $. Comparing (\ref{POLKERR}) and
(\ref{POLTH}), it is clearly seen that the operator (\ref{NLTH}) $
n_2 \left( \vec{E} \cdot \vec{E} \right)\vec{E}$ generalizes the
case of Marker and Terhune's operator (\ref{NLKERR}), but includes
also additional terms associated with THG.

The generalized nonlinear polarization (\ref{NLTH}) is quite simple
in the terms of left-hand and right-hand circularly components
\cite{BOYD,AGR}

\begin{eqnarray}
\label{ELF1} E_+(x,y,z,t) = A_+(x,y,z,t)\exp\left[ik_0(z-v_{ph}t)\right]
\nonumber\\
\\
E_-(x,y,z,t)= A_-(x,y,z,t)\exp\left[-ik_0(z-v_{ph}t)\right],\nonumber
\end{eqnarray}
where
\begin{eqnarray}
\label{CIRKKOMP} A_+=(A_x+iA_y)/\sqrt{2}\nonumber\\
\\
A_- =(A_x-iA_y)/\sqrt{2},\nonumber
\end{eqnarray}
are the left-hand $A_+$ and right-hand $A_-$ circular components of
the amplitude fields. Using the technique of calculations, presented
in \cite{BOYD}, we obtain that the nonlinear polarization
(\ref{NLTH}) generates the following components

\begin{eqnarray}
\label{NLCIRK} P_+ =
n_2\left(A_+^2A_-\right)\exp\left[ik_0(z-v_{ph}t)\right]\nonumber\\
\\
P_- = n_2
\left(A_-^2A_+\right)\exp\left[-ik_0(z-v_{ph}t)\right]\nonumber,
\end{eqnarray}
where $n_2$ is the nonlinear refractive index.

\section{Basic System of Equations}
The dynamics of narrow-band laser pulses can be accurately described
in the frame of paraxial optics. The filamentation experiments
demonstrate a typical pulse evolution: the initial laser pulse
$\left( t_0\geq 50 fs \right) $ possesses a relatively narrow-band
spectrum $\left(\Delta k_z \ll k_0 \right)$ and during the process
the initial self-focusing and self-compression broadens the spectrum
significantly. The broad-band spectrum $\left(\Delta k_z \sim
k_0\right)$ is one of the basic characteristics of the stable
filament. The evolution of the  filament can not be further
described in the frame of the nonlinear paraxial optics because the
paraxial optics works correctly for narrow-band laser pulses only.

Recently, the evolution of broad-band laser pulses was described
correctly within different non-paraxial models  such as UPPE
\cite{KOL1} or non-paraxial envelope equations \cite{KOVBOOK}.
Another standard restriction in the filamentation theory is the use
of one-component scalar approximation of the electrical field
$\vec{E}$. This approximation, though,  is in contradiction with
recent experimental results, where rotation of the polarization
vector is observed \cite{KOZA}. For this reason in the present paper
we use the non-paraxial vector model up to second order of
dispersion, in which the nonlinear effects are described by the
nonlinear polarization components (\ref{NLCIRK}). The system
non-paraxial equations of the amplitude functions $A_+$ and $A_-$
has the form

\begin{eqnarray}
\label{SYSCIRK1} -2ik_0 \left( \frac{\partial A_+}{\partial
z}+\frac{1}{v_{gr}}\frac{\partial A_+}{\partial t}\right)=
 \Delta A_+ - \frac{1+\beta}{v_{gr}^2}\frac{\partial^2 A_+}{\partial t^2}\nonumber\\
+ k^2_0 \tilde{n}_2A^2_+A_-\nonumber\\
\\
2ik_0 \left( \frac{\partial A_-}{\partial
z}+\frac{1}{v_{gr}}\frac{\partial A_-}{\partial t}\right)=
 \Delta A_- - \frac{1+\beta}{v_{gr}^2}\frac{\partial^2 A_-}{\partial t^2}\nonumber\\
+ k^2_0 \tilde{n}_2A^2_-A_+.\nonumber
\end{eqnarray}
where  $\Delta$ is $3D$ - $(x,y,z)$ Laplace operator,  $v_{gr}$ is the group  velocity, $\beta = k_0v_{gr}^2k''$ and
$k''$ is the group velocity dispersion. In the process of deriving
of the system of Eqs. (\ref{SYSCIRK1}) is used in mind, that the
left-hand and right-hand components are complex conjugated fields.

This model describes the ionization-free filamentation regime and we
will investigate the case, when the pulse intensities are slightly
above  to the critical one for self-focusing. The dispersion number
in air is too small $\beta = k_0v_{gr}^2k''\simeq 2.1\times10^{-5}$
and we will use Eqs. (\ref{SYSCIRK1}) in approximation of  the first
order of dispersion.

\begin{eqnarray}
\label{SYSCIRK} -2ik_0 \left( \frac{\partial A_+}{\partial
z}+\frac{1}{v_{gr}}\frac{\partial A_+}{\partial t}\right)=
 \Delta A_+ - \frac{1}{v_{gr}^2}\frac{\partial^2 A_+}{\partial t^2}\nonumber\\
+ \gamma A^2_+A_-\nonumber\\
\\
2ik_0 \left( \frac{\partial A_-}{\partial
z}+\frac{1}{v_{gr}}\frac{\partial A_-}{\partial t}\right)=
 \Delta A_- - \frac{1}{v_{gr}^2}\frac{\partial^2 A_-}{\partial t^2}\nonumber\\
+ \gamma A^2_-A_+,\nonumber
\end{eqnarray}
where $\gamma=n_2k_0^2A_0^2$ is nonlinear coefficient.

\section{Vector soliton solution and vector rotation. The filament as Rogue wave}

The nonlinear system of equation (\ref{SYSCIRK}) admits exact
solitary vector solution, when $\gamma=2$ and the spectral width of
the pulses reach the value $\Delta k_z\approx k_0$

\begin{eqnarray}
\label{soliton1}
A_+(x,y,z,t)=
\frac{2}{1+\tilde{r}^2}\exp\left[-i\Delta k_z\left(z-v_{gr}t\right)\right]\nonumber\\
\\
A_-(x,y,z,t)=\frac{2}{1+\tilde{r}^2}\exp\left[i\Delta
k_z\left(z-v_{gr}t\right)\right],\nonumber
\end{eqnarray}
where
$\tilde{r}=\sqrt{x^2+y^2+(z-ik_{cep})^2-v_{gr}^2(t-ik_{cep}/v_{gr})^2}$
and $k_{cep}$ is determinate below. The solution of the
corresponding vector electrical field can be written after
multiplying the amplitude functions (\ref{soliton1}) with the main
phase (\ref{ELF})

\begin{eqnarray}
\label{Efield}
E_+(x,y,z,t)=\frac{2}{1+\tilde{r}^2}
\exp\left[-i\Delta k_z\left(v_{ph}-v_{gr}\right)t\right]\nonumber\\
\\
E_-(x,y,z,t)=\frac{2}{1+\tilde{r}^2}\exp\left[i\Delta k_z\left(v_{ph}-v_{gr}\right)t\right].\nonumber
\end{eqnarray}
Lets turn from the left-hand and right-hand circular components
(\ref{CIRKKOMP}) to the standard Cartesian coordinates

\begin{eqnarray}
\label{CART} E_x = (E_+ +E_-)/\sqrt{2}, \; E_y =
(E_+-E_-)/(i\sqrt{2}).
\end{eqnarray}

\begin{figure*}
\centerline{\includegraphics[width=160mm,height=70mm]{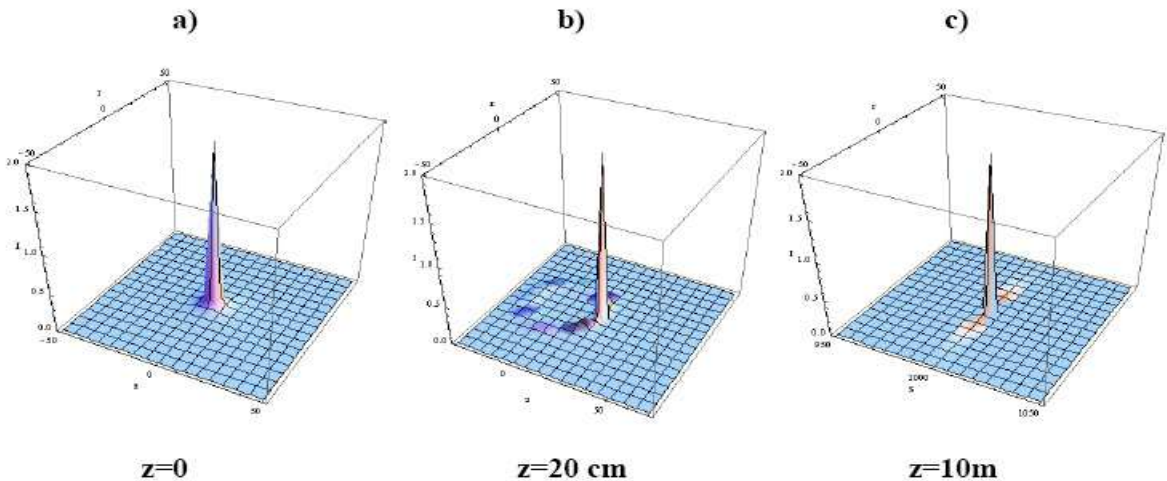}}
\caption{ Evolution of the intensity profile
$I=\left|A_x\right|^2+\left|A_y\right|^2$ of the solution
(\ref{REfield}). Side ($x,z$) projection is plotted. On Fig.1a) is
the initial profile, while on Fig.1b) the evolution of the intensity
profile on distance z=20 cm is presented. It is clearly seen the
generation of weak spherical wave. At large distances the solution
is relatively stable, as it can be seen from the Fig.1c). One more
detailed picture of the propagation in the far-away zone demonstrate
a weak self-compression of the pulse.}
\end{figure*}

The solution (\ref{Efield}) written in Cartesian coordinates has the
form

\begin{eqnarray}
\label{REfield}
E_x(x,y,z,t)=\frac{2}{1+\tilde{r}^2}\cos\left[\Delta k_z\left(v_{ph}-v_{gr}\right)t\right]\nonumber\\
\\
E_y(x,y,z,t)=\frac{2}{1+\tilde{r}^2}\sin\left[\Delta k_z\left(v_{ph}-v_{gr}\right)t\right].\nonumber
\end{eqnarray}
The $3D+1$ Lorentz type solution (\ref{REfield}), presented in
Cartesian coordinates,  gives  oscillation of the electrical vector
$\vec{E}=(E_x,E_y,0)$ in ($x,y$) plane. It can be seen directly that
the frequency of oscillation is equal to the carrier to envelope
ones $\omega_{cep}=k_0\left(v_{ph}-v_{gr}\right)$. The corresponding
longitudinal spatial carrier to envelope wave number is
$k_{cep}=\omega_{cep}/v_{gr}$.  In standard atmosphere this complies
to a rotation of the electrical field vector  with period in $z$
direction $\lambda_{cep}\cong 10-20$ $cm$. Other important
characteristic of the  vector solution (\ref{REfield}) is that it is
soliton in \emph{$3+1$ dimensional complex Minkowski space} and his
real dynamics in 3D space is more close to\emph{ weak Rogue wave}
with small self-compressing in the direction of propagation. The
evolution of the intensity profile is plotted on Fig.1 and
demonstrate a generation of  week spherical wave (Fig 1. b) in the
first milliseconds of the propagation distance. At large distances,
as it can be seen from the Fig. 1c), the propagation is relatively
stable with preserving the amplitude maximum. One more detailed
picture of the propagation in the far-away zone demonstrate a weak
self-compression of the pulse.

\section{Conclusions}
In this investigation we point that in the femtosecond region it is
impossible to reduce the nonlinearity of third order to the scalar
Kerr type only (proportional to the intensity). Therefore, we study
more precisely the vector character of the nonlinear third order
polarization, taking into account the carrier to envelope phase. The
corresponding amplitude equations are written in terms of left-hand
and right-hand circularly polarizing components. The vector system
(\ref{SYSCIRK}) admits exact (3+1)D soliton solutions
(\ref{soliton1}) of Lorentz type. Our soliton solution is obtained
for broad-band pulses $\Delta k_z\approx k_0$ and leads to the
conclusion that the soliton appears as a balance between parabolic
divergent type wave diffraction and parabolic convergent
nonlinearity. The solution gives also a rotation of the vector of
the electrical field with period equal to the period of carrier to
envelope wavelength $\lambda_{cep} \cong 10-20$ $cm$, in depends from the air pressure and temperature. Other
important result is the generation of weak spherical wave in the
first milliseconds of propagation and weak self compression in the
far-away zone.


\begin{thebibliography}{100}

\bibitem {MOU} A. Braun, G. Korn, X. Liu, D. Du, J. Squier, and G.
Mourou,  Opt. Lett. \textbf{20}, 73-75 (1995).

\bibitem{WO} L. W\"{o}ste, C. Wedekind, H. Wille, P. Rairoux,
B. Stein, S. Nikolov, C. Werner, S.Nierdermeier, F. Ronneberger, H.
Schillinger, and R. Sauerbrey, AT-Fachverlag, Stuttgard, Laser and Optoelectronik \textbf{29},
51-53 (1997).

\bibitem{TZ} S. Tzortzakis, G. M\'{e}chain, G. Patalano, Y.-B. Andr\'{e},
B. Prade, M. Franco, A. Mysyrowicz, J. M. Munier, M. Gheudin, G.
Beaudin, and P. Encrenaz, Opt. Lett. \textbf{27}, 1944-1946, (2002).

\bibitem{DAMYZ} C. D'Amico, A. Houard, M. Franco,  B. Prade,  A.
Mysyrowicz,  Optics Express \textbf{15}, 15274-15279 (2007).

\bibitem{HAU} C. P. Hauri, A. Guandalini, P. Eckle, W. Kornelis, J.
Biegert, U. Keller,   Optics Express \textbf{13}, 7541 (2005).

\bibitem{CHIN1} S. L. Chin, A. Brodeur, S. Petit, O. G. Kosareva, V.
P. Kandidov,  J. Nonlinear Opt. Phys. Mater. \textbf{8}, 121-146
(1999).

\bibitem{KOZA} O.Kosareva et al.,  Opt. Lett. \textbf{35}, 2904-2906 (2010).

\bibitem{KOPR} I. G. Koprinkov, A. Suda, P. Wang and K. Midorikawa, Phys. Rev. Lett. \textbf{84}, 3847 (2000).

\bibitem{KAN} S. L. Chin, S. A. Hosseini, W. Liu, Q. Luo, F. Th\'{e}berge, N.
Ak\"{o}zbek, A. Becker, V. P. Kandidov, O. G. Kosareva, and H.
Schoeder,  Can. J. Phys. \textbf{83}, 863-905 (2005).

\bibitem{KASP2} P. B\'{e}jot, J. Kasparian, S. Henin, V. Loriot,
T. Viellard, E. Hertz, O. Faucher, B. Lavorel, and J.-P.
Wolf, Phys. Rev. Lett., \textbf{104}, 103903 (2010).

\bibitem{MECH} G. M\'{e}chain,  A. Couairon, Y.-B. Andr\'{e}, C. D'Amico, M. Franco, B. Prade, S.
Tzortzakis,    A. Mysyrowicz, R. Sauerbrey,  Appl.Phys. B \textbf{79}, 379-382
(2004).
\bibitem{MECH2}  G. M\'{e}chain, \emph{Study of filamentation of femtosecond
laser pulses in air},  These de doctorat, Ecole Polytechnique,
Palaiseau, France, 2005.

\bibitem{KILO} Magali Durand et al.,  Optics Express, \textbf{21}, 26836 (2013).

\bibitem {MAKER} P. D. Maker and  R. W. Terhune, Phys. Rev. \textbf{137}, A801  (1965).
\bibitem{KOL1} M. Kolesik, J. V. Moloney,  Optics Express, \textbf{16}, 2971 (2008).
\bibitem{KOL2} M. Kolesik,  E. M. Wright, A. Becker, J. V. Moloney,   Appl. Phys. B, 85, pp 531-538 (2006).
\bibitem{LMK1} L. M. Kovachev, Journal of Modern Optics, \textbf{56}, 1797- 1803 (2009).
\bibitem{KOVBOOK} Lubomir M. Kovachev and Kamen Kovachev,  Laser System for Applications, Chapter 11, InTech, 2011.
\bibitem {BOYD}R. W. Boyd \textit{Nonlinear Optics}, ( 3th ed., Academic
Press, 2003).
\bibitem{AGR} G. P. Agrawal, \emph{Nonlinear Fiber Optics}, (4th ed.
Academic, 2007).

\end{thebibliography}
\end{document}